**Distances and peculiar velocities of spiral galaxies in the 2MFGC and SFI++ samples**

Yu.N.Kudrya[1], V.E. Karachentseva[1], I.D. Karachentsev[2], S.N.Mitronova[2], W.K.Huchtmeier[3]

[1] Astronomical observatory of the Kiev Taras Shevchenko National University, Observatorna 3, 04053, Kiev, Ukraine, e-mail: kudrya@observ.univ.kiev.ua.
[2] Special Astrophysical Observatory, Russian Academy of Sciences, N.Arkhyz, KChR, 369167, Russia, e-mail: ikar@luna.sao.ru
[3] Max-Planck-Institut für Radioastronomie, Auf dem Hügel 69, 53121, Bonn, Germany.

Abstract

We compare infrared Tully-Fisher (TF) distances and peculiar velocities derived for spiral galaxies from the two largest datasets: the 2MASS selected Flat Galaxy Catalog, 2MFGC [19, 20] and the Arecibo General Catalog with I-band photometry, SFI++ [30,7]. These samples contain peculiar velocities for ~3000 and ~4000 objects, respectively. Based on a sub-sample of ~1000 common deeply inclined galaxies, we reach the following conclusions. Irrespective to high (SFI++) or low (2MFGC) quality of the used photometric data, about 10% of the galaxies in both samples deviate considerably from the main body of the TF relation. After their deletion, the standard TF scatters drops to $0.47^m$ (2MFGC) and $0.40^m$ (SFI++). The TF distances, derived from two the samples, demonstrate a high degree of mutual agreement with a correlation coefficient $\rho = +0.95$ and $\sigma(H_0 r) = 837$ km/s. Peculiar velocities of the galaxies are also correlated with $\rho = 0.56 - 0.59$ and $\sigma(V_{pec}) = 610$ km/s. We find that the bulk motion of the 2MFGC and SFI++ galaxies on a typical scale of $H_0 r \approx 5700$ km/s can be represented by a dipole solution with the amplitude $V = (297 \pm 23)$ km/s directed towards $\{l = 292^0 \pm 4^0, b = -12^0 \pm 3^0\}$, being only slightly sensitive to different modifications of the TF relaton.

Резюме

Мы сравниваем инфракрасные Талли-Фишер (TF) расстояния и пекулярные скорости, определенные для спиральных галактик из двух наибольших выборок: 2MASS каталога плоских галактик, 2MFGC [19, 20] и The Arecibo General Catalog с фотометрией в I-полосе, SFI++ [30, 7]. Эти выборки содержат пекулярные скорости для ~3000 и ~4000 объектов соответственно. Основываясь на подвыборке ~ 1000 общих сильно наклоненных галактик, мы пришли к следующим выводам. Независимо от высокого (SFI++) и низкого (2MFGC) качества фотометрических данных, в обеих выборках доля галактик, которые сильно отклоняются от TF зависимости, составляет ~10%. После их исключения стандартные TF отклонения уменьшаются до $0.47^m$ (2MFGC) и $0.40^m$ (SFI++). TF-расстояния, определенные для этих двух выборок, показывают высокую степень взаимного согласия с коэффициентом корреляции $\rho = +0.95$ и стандартным отклонением $\sigma(H_0 r) = 837$ км/с. Пекулярные скорости галактик также скоррелированы с $\rho = 0.56 - 0.59$ и $\sigma(V_{pec}) = 610$ км/с. Мы нашли, что параметры коллективного движения 2MFGC и SFI++ галактик мало чувствительны к различным модификациям TF зависимости. На типичном масштабе $H_0 r \approx 5700$ дипольное решение для обеих выборок характеризуется амплитудой $V = (297 \pm 23)$ км/с и направлением $\{l = 292^0 \pm 4^0, b = -12^0 \pm 3^0\}$.

**1 Introduction**

The Tully-Fisher (TF) relation [1] determines an empirical association between rotational velocity or width of the 21-cm radio line and absolute magnitude (luminosity) of a spiral galaxy. This diagram is widely used for deriving galaxy distances $H_0 r$ and peculiar velocities defined as $V_{pec} = cz - H_0 r$, where $cz$ is a measured radial velocity of a galaxy and $H_0$ is the Hubble constant. Within the context of the linear theory of gravitational instability, peculiar velocities of galaxies are related with mass fluctuations via cosmological parameters [2]. So, matter distribution in a sufficiently large volume can be restored from an observed field of peculiar velocities in it by giving a set of cosmological parameters and boundary conditions [3]. Analysis of peculiar velocities gives an opportunity to determine direction and modulus of the velocity of bulk motions of galaxies, and the distance at which the local stream of galaxies becomes convergent.

To solve this and other cosmological problems we need a catalog of peculiar velocities for an "ideal" sample of galaxies which would be characterized by a large number of galaxies, their uniform distribution over the sky, a sufficient redshift depth and a high quality of observational data. Such a single sample does not exist yet. During 30 years different groups of astronomers accumulated extensive observational datasets which were subsequently used to determine parameters of coherent motions of galaxies, deriving peculiar velocity field, calculation of cosmological parameters, etc. Data on different samples of galaxies show satisfactory mutual agreement of parameters of the bulk motion within a distance $H_0 r$ up to 5000 km/s, but at larger distances a discrepancy of results is pointed out [4,5]. Some lists of peculiar velocities were briefly compared in [6,7].

Scatter in the TF-diagram can be reduced considerably at transition from optical to infrared photometry which was first applied in [8,9,10] (see also a review adduced in the thesis [11]). An example of creation of a uniform sample for studying the large-scale non- Hubble flows is the Flat

Galaxy Catalog, FGC [12] and its updated version RFGC [13]. Spiral edge-on galaxies were successfully used for construction of the TF relation and estimation of distances [14]. Application of the simplest selection criterion of thin galaxies with an apparent axial ratio of $a/b > 7$ allowed us selecting disc-dominated (Sc-Sd) galaxies of the general field uniformly distributed over all the sky. The RFGC catalog contains 4236 galaxies. Rotational velocities or the 21-cm radio line widths are known for about 1600 of them. By these data and by optical parameters adduced in RFGC, a list of individual peculiar velocities and distances for 1327 galaxies was compiled basing on the TF relation in its "linear diameter – line width" version [15]. Successful application of the infrared TF relation to RFGC galaxies [16] suggested us a possibility of selecting flattened spiral galaxies from the Extended Sources Catalog (XSC) [17,18] on a basis of their 2MASS characteristics. The 2MASS selected Flat Galaxy Catalog 2MFGC [19] contains 18020 galaxies with an infrared (NIR) axial ratio of $b/a < 0.3$ that corresponds approximately to an optical axial ratio of $a/b > 6$. Among them there are 3074 spiral galaxies with measured characteristics for constructing the TF relation. Comparison of distances calculated independently from optical and infrared data reveals their good mutual agreement [20]. A coefficient of correlation between optical and NIR distances turns out to be about +0.95. Parameters of the dipole solution for the peculiar velocity fields at a characteristic scale of ~6000 km/s turn out to be close to each other also.

Recently, Springob et al. [7] published the SFI++ catalog which holds observational data on the TF-diagram and also distances and peculiar velocities for 4053 spiral galaxies in clusters and in the field. When compiling this catalog, they make use of digital optical images, optical long-slit spectra, and global HI line profiles to extract parameters of relevance to disk scaling relations, incorporating several previously published datasets as well as a new photometric sample of some 2000 objects. In this paper we intend to find out how estimates of TF-distances and peculiar velocities of spiral galaxies depend on properties of an initial sample, as well as a method of correction of observables and a manner of data analysis. With this idea in mind, we compare distances and peculiar velocities held in the catalogs [20, 7].

**2 Determination of distances and peculiar velocities by the Tully- Fisher method**

The sky distribution of 3074 galaxies from the 2MFGC sample and 4053 galaxies from the SFI++ sample are shown in galactic coordinates $\{l,b\}$ in the upper and lower panels of Fig.1, respectively. As it is seen from their comparison, galaxies from the infrared catalog 2MFGC show a more uniform distribution and better fill the Milky Way region.

### 2.1. The 2MFGC sample

The data sources about magnitudes, radial velocities, HI widthes and rotational velocities have been described in detals in the articles [21, 22]. Distances and peculiar velocities were determined by us in two stages. At the first step we calibrated a linear multiparametric TF relation of the following form:

$$M = c_0 + c_1 \cdot \log W_{50}^c + c_2 \cdot \log(a/b) + c_3 \cdot Jhl + c_4 \cdot (J_{fe}^c - K_{fe}^c) + c_5 \cdot Jcdex \quad (1)$$

where $M$ denotes galaxy absolute magnitude, and $W_{50}^c$ is the HI line width corrected for cosmological widening. The following photometric characteristics were taken from the XSC catalog: the Kron "elliptic" $J_{fe}$- and $K_{fe}$- magnitudes, the $J$ - band effective surface brightness $Jhl$, the concentration index $Jcdex$ (ratio of radii inside which 3/4 and 1/4 of galaxy light is concentrated), and the major-to-minor axial ratio $a/b$. The Kron $J$ - and $K$ - magnitudes were corrected for the Galactic extinction according to [23]. The term with $\log(a/b)$ in (1) takes account of an internal extinction in the galaxy, and other terms trace a structural proportion between the galaxy bulge and disk.

When making relations (1), we assumed an arbitrary number of regressors in the right-hand side which reflect the galaxy structure but do not depend (like $W_{50}$) on the galaxy distance. Sequential analysis showed that only five of them, kept in (1), significantly reduce a scatter in the TF diagram. The absolute magnitude of a galaxy in the left-hand side of (1) was calculated by its apparent magnitude $J_{fe}^c$ as:

$$M = J_{fe}^c - 25 - 5\log r, \quad (2)$$

where the photometric distance $r$ (in Mpc) was expressed via radial velocity by the post-Hubble relation

$$r = V_{3K}\{1 - (q_0 - 1) \cdot V_{3K}/2c\}/H_0, \quad (3)$$

which is valid for the uniform isotropic cosmological model. Here $c$ is the velocity of light. The radial velocity $V_{3K}$ with respect to the cosmic microwave background (3K) was calculated by its measured heliocentric radial velocity $V_h \equiv cz$ with the use of parameters of the Sun motion relative to the cosmic microwave background obtained in [24]. The Hubble constant was accepted to be $H_0 = 75$ km/s/Mpc. The value of deceleration parameter $q_0 = -0.55$ was taken correspondent to the standard flat cosmological model with cold dark matter and the cosmological constant $\Omega_\Lambda = 0.7$. Note that application of (3) instead of the linear Hubble relation increases the distance estimate not more than by 2.5% for the most distant galaxies in the 2MFGC sample.

The calibration constants $c_i$ in (1) are calculated by minimizing the sum of squared deviations of the right-hand sides in (1) and (2) with account of (3). Using the calculation scheme (1)- (3), we had to ignore at this stage the peculiar velocities which are especially essential for the nearest galaxies. Since we minimize the sum of squared deviations, a slight asymmetry called the Malmquist bias arises because of the term $\lg(V \pm \Delta V)$. It manifests itself in "swelling" of the TF diagram which is the largest for nearby dwarf galaxies.

The TF diagram for the sample of 3074 2MFGC galaxies is given in the upper left part of Fig.2. Its scatter is rather large: $\sigma_{TF} = 0.^m76$. Following [20], we eliminated galaxies deviating from the regression line by more than $3\sigma_{TF}$, and also galaxies with peculiar velocities in the $3K$ reference frame higher than 3000 km/s, assuming that such deviations are caused by observational errors, but not physical reasons. After making several consecutive eliminating iterations (up to the process convergence) we obtained a cleaned sample of N=2724 which is characterized by the scatter $\sigma_{TF} = 0.^m47$. The TF relation for the cleaned sample is given in the upper right part of Fig.2.

At the second step, after calibrating the TF relation, we determined individual galaxy distances $H_0 r$ from (1) and (2), and calculated their individual peculiar velocities as

$$V_{pec} = V_{3K} - H_0 r \{1 + (q_0 - 1)H_0 r / 2c\}. \qquad (4)$$

As a result, individual distances and peculiar velocities of 2724 2MFGC galaxies were presented in [20].

**2.2. The SFI++ sample**

This sample was compiled as a sum of the following subsamples.

a) The SCI (24 clusters having mean velocities less than 10000 km/s).

b) The SC2 (52 all sky clusters with recessional velocities $5000 < cz < 25000$ km/s).

c) The SFI was comprised of a TF sample of 2000 field galaxies limited to $cz < 7500$ km/s, blue magnitude $m_B < 14.5$, and line width $W_{50}^c > 100$ km/s.

d) The SF2 program was intended to obtain photometry for objects either with existing HI or optical spectroscopy in existing database and to target the region $-15^0 < \delta < +35^0$ to a depth of $cz < 10000$ km/s and optical diameter $a > 0.9'$.

e) The Mathewson, Ford, & Buchhorn [25] dataset was a compilation of TF measurements for 1355 Sb-Sd galaxies in the southern hemisphere, diameters $a > 1.7''$, inclinations $> 40^o$, and Galactic latitude $|b| > 11^0$. Most of the objects had systemic velocities of less than 7000 km/s.

f) The Mathewson & Ford [26] dataset was a compilation of TF measurements for an additional 920 Sb-Sc galaxies selected from the ESO-Uppsala Survey of the ESO(B) Atlas [27], with diameters $1.0' < a < 1.6'$, systemic velocities between 4000 and 14000 km/s.

g) An additional 172 Uppsala General Catalog [28] galaxies were observed in the region $250^0 < l < 360^0$, $45^0 < b < 80^0$. Both the photometric and optical spectroscopic datasets from [25, 26] have been reprocessed using the algorithms [29] to achieve greater homogeneity.

Data from columns 6 and 10 of Table 2 in [7] were used for making the TF diagram. The Tully-Fisher relation for the sample of 4053 SFI++ galaxies is represented in the left lower panel of Figure 2. We cleaned this sample by the manner accepted by us for 2MFGC. The cleaned TF diagram for 3695 SFI++ galaxies is shown in the right lower part of Figure 2. Elimination of SFI++ galaxies with deviations exceeding $3\sigma$ and peculiar velocities higher than 3000 km/s reduces a scatter in the TF diagram from $\sigma_{TF} = 0.^m57$ to $0.^m39$. Comparison of the top and bottom panels in Fig.2 shows a lower scatter of the SFI++ galaxies than the 2MFGC ones. This difference is caused by much higher quality of photometric data in the SFI++ sample and by different manner of estimating the absolute magnitude of nearby dwarf galaxies with their relatively larger errors in luminosity because of peculiar velocities.

## 3 Common galaxies in the 2MFGC and SFI++ samples

### 3.1. Comparison of initial data.

Comparison of radial velocities for 983 galaxies which are common in the two samples shows that these arrays are practically identical. Considerable differences are found for four objects only (2MFGC 3827, 9300, 15222, 17898; respective Arecibo numbers 23663, 211029, 32909, 330958). Figure 3 represents comparison of the 21-cm line width for galaxies common in the SFI++ and 2MFGC samples. The line widths measured at the 50% level of peak were corrected for the galaxy disk incline and cosmological widening. In the SFI++ sample the authors also introduced a correction for turbulent motions in galaxies. The solid line in Fig.3 represents the diagonal, and the dashed lines are shifted up and down from the diagonal by $3\sigma = 0.098$. Galaxies outside the band (N=15) are denoted by their numbers in 2MFGC. On the whole, differences in the data arrays are not large. In some cases (2MFGC 1249, 1685, 17422, 17786) the width differences are caused by different estimates of galaxy incline angle derived from *J*- or *I*-band isophots. Note, that in spite of some differences in the linewidth correction, we find $\sigma(\Delta \lg W^c) = 0.03$, i.e. the rms scatter in the corrected linewidths turns out to be 8% only. However, with the TF relation slope of about -7, this equals to the magnitude scatter of $\sigma_{TF} \sim 0.^m 2$, which is comparable to a photometric error $\sim 0.^m 25$ in the 2MASS sample.

Figure 4 demonstrates the relation between apparent magnitudes of 983 common galaxies. The *I*-band magnitudes corrected for internal and external (Galactic) extinction are taken from [7], and *J*-band magnitudes corrected only for the Galactic extinction [23] are taken from 2MASS. The regression line $m_I = m_J + 0.412$ is shown in the Figure. The galaxy scatter relative to it is $\sigma_m \sim 0.^m 28$. Eleven galaxies with deviations from the straight line exceeding 3σ are denoted by their 2MFGC numbers. The distribution of 11 deviating galaxies looks asymmetrical. It is caused mostly by underestimating total luminosities of galaxies with faint extended periphery based on the shallow 2MASS photometry.

**3.2. Comparison of distances**. Panels of Figure 5 reproduce comparison of distances determined for 983 objects common in the 2MFGC catalog and the SFI++ sample without (left panel) and with (right panel) allowance made for the Malmquist bias. The dashed line denotes the line of direct regression, and the solid line is the diagonal (in the right panel it is practically indistinguishable from the regression line). From the adduced analysis it follows that the 2MFGC distances are in a bit better agreement with the SFI++ distances obtained with allowance made for the Malmquist bias. Since we determined the 2MFGC distances on a basis of multiparameter generalization of the TF relation, it can be assumed that the use of additional regressors reduces the Malmquist bias, at least, partially. We note also that in the case of orthogonal regression with

$$H_0 r (SFI++) = (1.024 \pm 0.005) \cdot H_0 r (2MFGC)$$

the Fisher statistics is equal to 1.8. Comparing this value with the Fisher quantile for the 95% significance level (3.84), we conclude that this hypothesis may be statistically consistent. In such a case, we have the straight proportion of distances with the rms scatter $\sigma(H_0 r) = 840$ km/s, and $\rho = +0.95$.

**3.3. Comparison of peculiar velocities**

Two panels in Figure 6 presents comparison of peculiar velocities for 983 common galaxies. Ellipses in the Figure are boundaries of the 95%-probability level, if one accepts the plane distribution of points to be a two-dimensional Gaussian one with averages, dispersions, and a correlation coefficient ρ determined from the observational data. Our calculations yield $\rho = 0.56$, $\sigma = 1110$ km/s for the left diagram and $\rho = 0.59$, $\sigma = 898$ km/s for the right one. This means that peculiar velocities in the 2MFGC sample are closer to peculiar velocities in the SFI++ sample with allowance made for the Malmquist bias, than to peculiar velocities without it.

The outermost dashed lines in Fig.6 correspond to the direct and inverse regressions, and the middle dashed line denotes the orthogonal regression. In the case of peculiar velocities with allowance made for the Malmquist bias, the orthogonal regression takes the form:

$$V_{pec}(SFI++) = (1.00 \pm 0.04) \cdot V_{pec}(2MFGC) - (230 \pm 28),$$

i.e. the velocities $V_{pec}(SFI++)$ are, on average, 230 km/s less than the peculiar velocities $V_{pec}(2MFGC)$.

## 4 Parameters of dipole solution

### 4.1. The 2MFGC sample.

Now we use arrays of peculiar velocities $V_{pec}$ to calculate orthogonal components $\vec{V} = (V_x, V_y, V_z)$ of a dipole term of bulk velocity

$$V_{pec,i} = \vec{V} \cdot \vec{e}_i + V_{p,i} \quad (5)$$

by minimizing the sum of squares of a "noise" component $V_{p,i}$ of the peculiar velocity ($i$ is the order number of a galaxy in the sample). Here $\vec{e}_i = (\cos l_i \cos b_i, \sin l_i \cos b_i, \sin b_i)$ is a unit vector of direction to galaxy $i$ in the reference related to galactic coordinates $l, b$. Then we calculate the module $V_B$ and direction $l_B, b_B$ of the bulk motion of galaxies by orthogonal components.

In Table 1 we present for the 2MFGC sample results of calculation of parameters of the dipole solution together with the slope coefficient $c_1$ of the TF relation (simple or generalized), the rms deviation $\sigma_{TF}$ from the TF relation, the rms noise component of the peculiar velocity $\sigma_V$ and the Fisher significance $F$ of the vector of bulk motion. The fourth column refers to the formula number of the used regression. The first two rows of Table 1 present results of calculation with the use of the multiparametric TF relation (1) and eqs. (2) and (3) for the complete (N=3074) and cleaned (cln) (N=2724) galaxy samples. Recall that the cleaning of data is a sequential deletion of galaxies which either give deviations exceeding $3\sigma$ in the TF diagram or lead to unfeasible peculiar velocities exceeding modulo 3000 km/s.

Table 1: Parameters of the TF relation and dipole solution for the 2MFGC galaxy sample.

| n | Sample | N | regr. | $\sigma_{TF}$ | $c_1$ | $\sigma_V$, km/s | $V_B$, km/s | $l_B$, deg | $b_B$, deg | F |
|---|--------|---|-------|---------------|-------|------------------|-------------|------------|------------|---|
| 1 | 2MFGC | 3074 | (1) | 0.763 | -4.66±0.10 | 1668 | 169±56 | 314±19 | -7±15 | 3.1 |
| 2 | 2MFGC(cln) | 2724 | (1) | 0.471 | -6.53±0.08 | 1018 | 199±37 | 304±11 | -8±8 | 9.8 |
| 3 | 2MFGC | 3074 | (6) | 0.874 | -6.01±0.09 | 1892 | 411±65 | 314±9 | -32±7 | 14 |
| 4 | 2MFGC(cln) | 2676 | (6) | 0.482 | -7.53±0.06 | 1050 | 313±39 | 304±7 | -20±5 | 21 |
| 5 | 2MFGC | 3074 | (7) | - | -9.07 | 2073 | 463±72 | 300±9 | -32±7 | 14 |
| 6 | 2MFGC(cln) | 2663 | (7) | - | -9.07 | 1091 | 334±41 | 291±7 | -19±5 | 22 |

Results of the second row correspond exactly to the dipole solution discussed by us earlier [20]. The coefficients $c_i$ in (1) for the cleaned sample together with their significance according to the Fisher criterion (given in parentheses) turn out to be:

$$c_0 = -9.95 \pm 0.43 \ (541), \quad c_1 = -6.53 \pm 0.08 \ (6696),$$

$$c_2 = 1.17 \pm 0.05 \ (466), \quad c_3 = 0.228 \pm 0.016 \ (208),$$

$$c_4 = -0.53 \pm 0.07 \ (53), \quad c_5 = 0.016 \pm 0.009 \ (3).$$

These quantities manifest different statistical significances of the regressors used in the multiparameter TF relation (1).

Then we introduce some variations in the calculation method for the same sample to estimate how much the value and direction of the dipole solution are dependent on the variations.

a) We repeated calculation using the simple TF relation instead of (1)

$$M - 5 \cdot \log(h) = c_0 + c_1 \cdot \log W_{50}^c, \tag{6}$$

where $h$ is the Hubble constant in units of 100 km/s/Mpc. The results are given in rows 3 and 4 of Table 1. Comparing these data with those indicated in the previous rows, we note that the rejection of additional regressors in the TF relation noticeably increases the dipole amplitude and shifts its direction toward lower galactic latitudes. In so doing, the procedure of cleaning excludes somewhat more galaxies (398 instead of 350), and the slope of the TF relation increases.

b) We derived dipole solutions without calculating coefficients of the Tully-Fisher relation by the 2MFGC data, but using the simple TF relation with parameters obtained in [30] for the SFI++ sample:

$$M_J - 5 \cdot \log(h) = -21.00 - 9.07 \cdot (\log W_{50}^c - 2.5). \tag{7}$$

The results are presented in last two rows of Table 1. Surprisingly, in spite of considerable increasing the slope of the TF relation, the dipole parameters turned out to be close to those obtained with the TF relation (6) calibrated directly by the 2MFGC data.

**4.2. The SFI++ sample.** Then we processed data on the SFI++ sample following the scheme accepted before in [20] and [6] with some modifications:

a) At first we derived the dipole solution using a simple TF relation with the parameters from [7]:

$$M_I - 5 \cdot \log(h) = -20.85 - 7.85 \cdot (\log W_{50}^c - 2.5). \tag{8}$$

The results are given in the first two rows of Table 2, where notations are similar to those of Table 1.

b) Then we accepted the TF relation (6) in the manner, which has been applied by us to the 2MFGC sample. Absolute magnitudes and line-width logarithms for the SFI++ sample were taken from Table 2 in [7]. The TF diagrams for the initial (N=4053) and cleaned (N=3695) samples are shown in lower panels of Fig.2. Results of calculation of parameters of the TF relation and the dipole bulk motion are given in the rows 3 and 4 of Table 2. As may be seen, deletion of only 9% of unreliable observational data from the initial sample increases the slope of the TF relation (from 6.65 to 7.12) and considerably decreases dispersion (from $0.^m57$ to $0.^m39$). In so doing, the dipole amplitude decreases noticeably, but its direction remains approximately unchanged.

Table 2. Parameters of the TF relation and dipole solution for the SFI++ galaxy sample.

| n | Sample | N | *regr.* | $\sigma_{TF}$ | $c_1$ | $\sigma_V$, km/s | $V_B$, km/s | $l_B$, deg | $b_B$, deg | $F$ |
|---|--------|---|---------|---------------|-------|------------------|-------------|------------|------------|-----|
| 1 | SFI++ | 4053 | (8) | - | -7.85 | 1669 | 358±51 | 291±9 | -3±6 | 18 |
| 2 | SFI++(cln) | 3674 | (8) | - | -7.85 | 1044 | 266±34 | 281±7 | -6±6 | 21 |
| 3 | SFI++ | 4053 | (6) | 0.568 | -6.65±0.06 | 1528 | 397±46 | 295±7 | -5±5 | 26 |
| 4 | SFI++(cln) | 3695 | (6) | 0.394 | -7.12±0.04 | 1012 | 320±33 | 288±6 | -11±4 | 32 |
| 5 | SFI++ | 3655 | (9) | 0.555 | -6.19±0.07 | 1338 | 431±42 | 294±6 | -1±5 | 42 |
| 6 | SFI++(cln) | 3405 | (9) | 0.407 | -6.87±0.06 | 977 | 348±34 | 287±6 | -6±4 | 40 |

c) Table 1 in [7] holds various galaxy parameters obtained on a basis of their $I$-band photometry: average surface brightness $SB$, ellipticity of galaxy shape $e$, ratio of isophotal radius to effective one, $r_{23.5}/r_{83L}$, and morphological galaxy type $T$ in digital code. All these parameters do not depend on distance (if neglecting the Tolmen effect for surface brightness) and can serve as any equivalents of additional regressors which we used in (1): $SB$ is an analog of the index $Jhl$, $(1-e)$ is an axial ratio, $T$ is an analog of the color index $(J-K)$, and $r_{23.5}/r_{83L}$ is an analog of the concentration index *Jcdex* characterizing a disk/bulge ratio. By analogy with (1) we used the multiparameter TF relation

$$M_I - 5 \cdot \log(h) = c_0 + c_1 \cdot \lg W_{corr} + c_2 \cdot \log(1-e) + c_3 \cdot \lg(r_{23.5}/r_{83L}) + c_4 \cdot T \quad (9)$$

applying it to the initial and cleaned SFI++ samples. A number of galaxies fell out of analysis because of lack of additional data for them. Results of our calculation are shown in two last rows of Table 2. The coefficients $c_i$ in the relation (9) for the cleaned sample together with their significance according to Fisher criterion (given in parentheses) turn out to be:

$$c_0 = -3.86 \pm 0.15 \ (661), \quad c_1 = -6.87 \pm 0.05 \ (13890),$$
$$c_2 = -0.421 \pm 0.038 \ (123), \quad c_3 = -0.63 \pm 0.10 \ (36),$$
$$c_4 = -0.0188 \pm 0.0054 \ (12).$$

It should be noted that the surface brightness SB is found to be a statistically insignificant regressor, and its deletion from (9) actually does not change parameters of the dipole solution. On the whole, introduction of additional regressors slightly deceases the TF relation slope and increases amplitude of bulk motions in the dipole approximation.

**5 Discussion and conclusions**

In this paper we tried to outline how much the estimates of distances and peculiar velocities of spiral galaxies determined by the Tully-Fisher method depend on features of forming an initial sample and a manner to analyse these data. We considered the two most nowadays-representative samples of spirals which have been compiled on essentially different principles. A number of galaxies that are common for both the samples is sufficiently large (N=983) for statistical comparison, but is a small part of the samples themselves. Again, the photometric data used for common galaxies are completely independent. The main results of the comparison can be formulated as follows.

Irrespective to quality of photometric dataset (deep in *I* band and shallow in the 2MASS survey) about 10% of spiral galaxies show considerable deviations from the main body of the TF relation. Reasons of these deviations can be very various: measurement errors in cases of noisy 21-cm line, uncertain correction for inclination for galaxies with asymmetric structure, unaccurate photometry for low surface brightness objects, confusion of HI signals from two or more galaxies situated within a telescope beam, etc. Most significant deviations fall on dwarf galaxies with their small amplitude of internal motions. Elimination of these 10% of low-quality observational data essentially reduces the dispersion $\sigma^2_{TF}$. The slope of the TF relation therewith increases appreciably.

For the sample of 983 common edge-on galaxies having the median depth of $H_0 r \approx 5700$ km/s their TF distances determined by independent methods show a high degree of agreement with the correlation coefficient $\rho = 0.95$ and the rms deviation $\sigma(H_0 r) = 837$ km/s. Herein the account or neglect of the Malmquist bias affects faintly the value $\rho$. The mutual agreement of the TF distances looks somewhat better with allowance made for the Malmquist bias in the SFI++ sample.

Residual non-Hubble (peculiar) velocities also demonstrate a considerable level of correlation with $\rho = 0.56 - 0.59$; here the second value corresponds to the case when the Malmquist bias was taken into account. Relation between the two datasets of $V_{pec}$ can be expressed by the linear orthogonal regression $V_{pec}(SFI++) = (1.00 \pm 0.04) \cdot V_{pec}(2MFGC) - (230 \pm 28)$ with the rms deviation 614 km/s.

As was repeatedly noted, one or another manner to determine TF distances and peculiar velocities for spiral galaxies is acted upon by a number of obvious and hidden selection effects (the Malmquist bias, dependence of the TF parameters on galaxy morphology and environment, probable curvature of the TF relation, ignoring or including in the consideration of dwarf galaxies, etc.). The integral checking for these veiled effects can be a reproducibility of parameters of the peculiar velocity field in the simplest, dipole approximation. Results of determining the amplitude and direction of the dipole component $\{V_B, l_B, b_B\}$ are presented in Tables 1 and 2, where both simple and multiparameter versions of the TF relation for both samples were used. From our analysis it follows that a modification of the TF regression form by means of allowing for the galaxy morphologic type, surface brightness, index of light concentration, and apparent axial ratio changes the apex location at the level $(1 \div 2)\sigma$, while amplitude of the bulk motion changes to a greater extent, at the level $(2 \div 3)\sigma$.

Taking into consideration the statistical "goodness" $G = (N/100)^{1/2} / \sigma_{TF}$ of two considered samples ($G = 11.1$ for 2MFGC and $G = 14.3$ for SFI++), we derive the weight-average dipole parameters: $V_B = (297 \pm 23)$ km/s, $l_B = 292^\circ \pm 4^\circ$, $b_B = -12^\circ \pm 3^\circ$. The obtained values are in a satisfactory fit with the estimates $\{V_B, l_B, b_B\}$ derived by MarkIII team [31] and some other samples, with due regard to their lower goodness. The observed agreement between directions and amplitudes of bulk motions for different samples of spiral galaxies on the scale $\sim 6000$ km/s allows us assuming that a combined sample of $N \sim 6000$ disk-dominated galaxies (with the expected goodness of $G \sim 20$) will serve as a firm ground for studying in details the local peculiar velocity field.

Acknowledgements

We wish to thank Martha Haynes for fruitful discussions. This work was partly supported by the DFG-RFBR grant 06-02-04017 and the national NASU programme "Cosmomicrophysics".

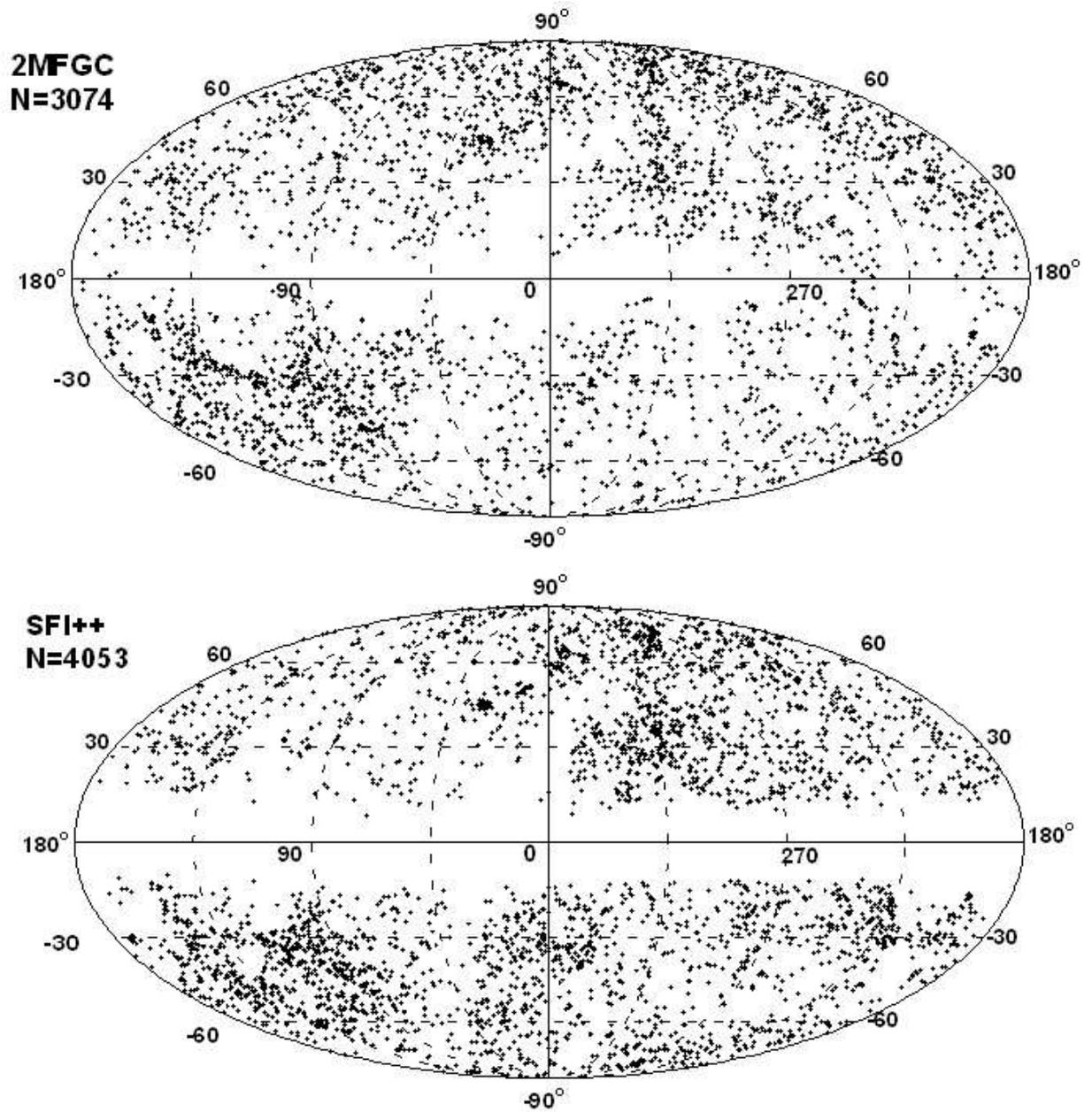

Figure 1: Sky distribution of galaxies from the 2MFGC sample (top panel) and the SFI++ sample (bottom panel) in galactic coordinates.

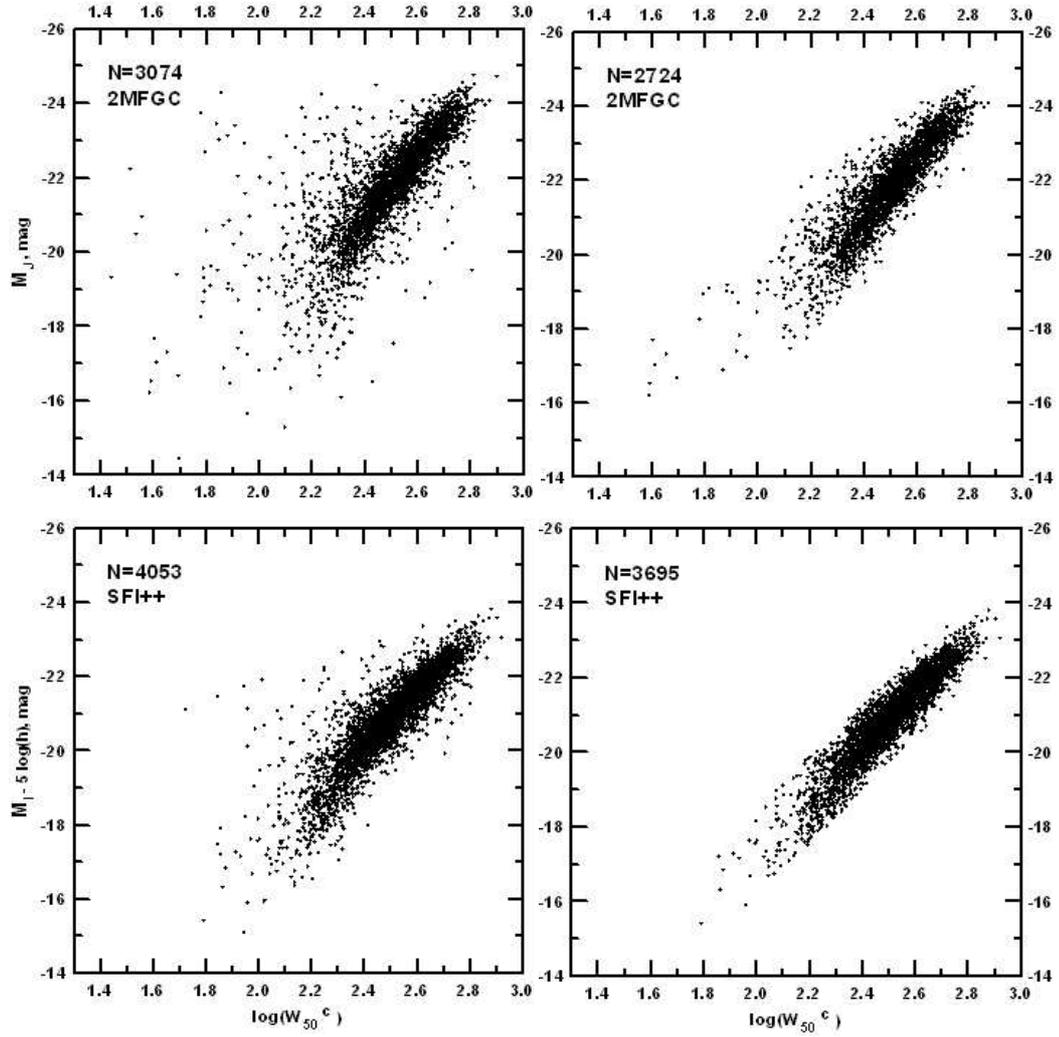

Figure 2: Top: The Tully-Fisher diagram for the 2MFGC sample before cleaning (left) and after it (right panel). Bottom: The TF diagram for the initial (left panel) and cleaned (right panel) sample of SFI++ galaxies.

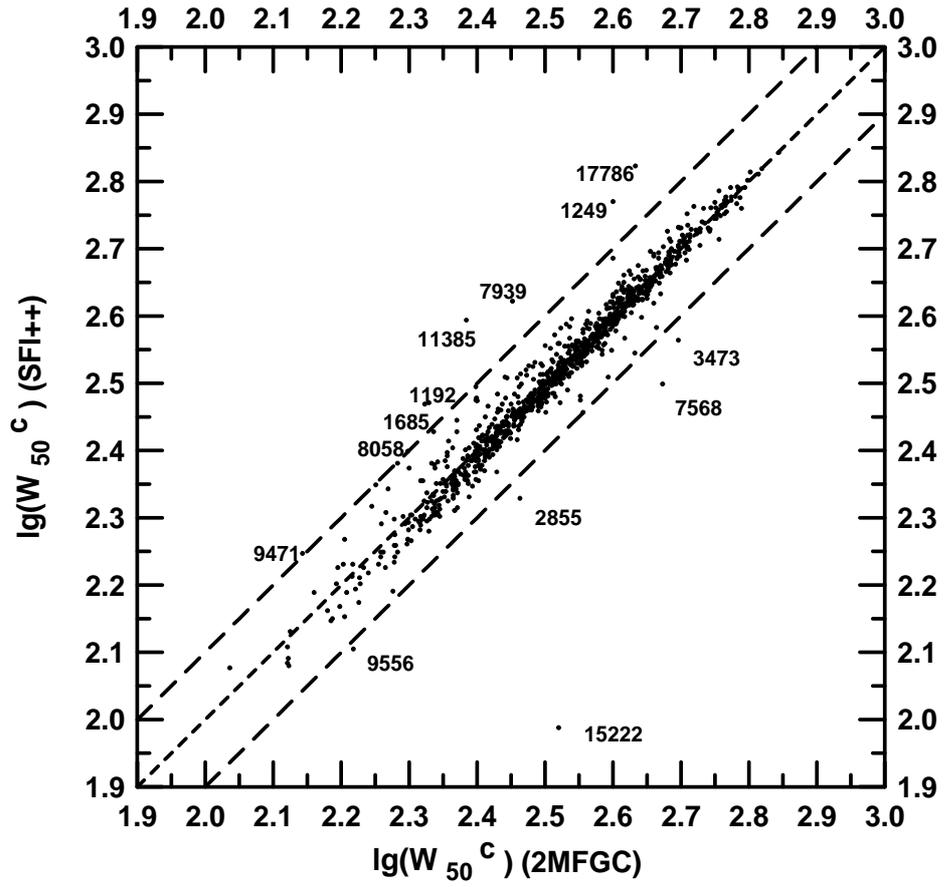

Figure 3: Comparison of corrected line-widths in the SFI++ and 2MFGC samples for 983 common edge-on galaxies. Galaxies outside the 3σ band are marked by their 2MFGC numbers.

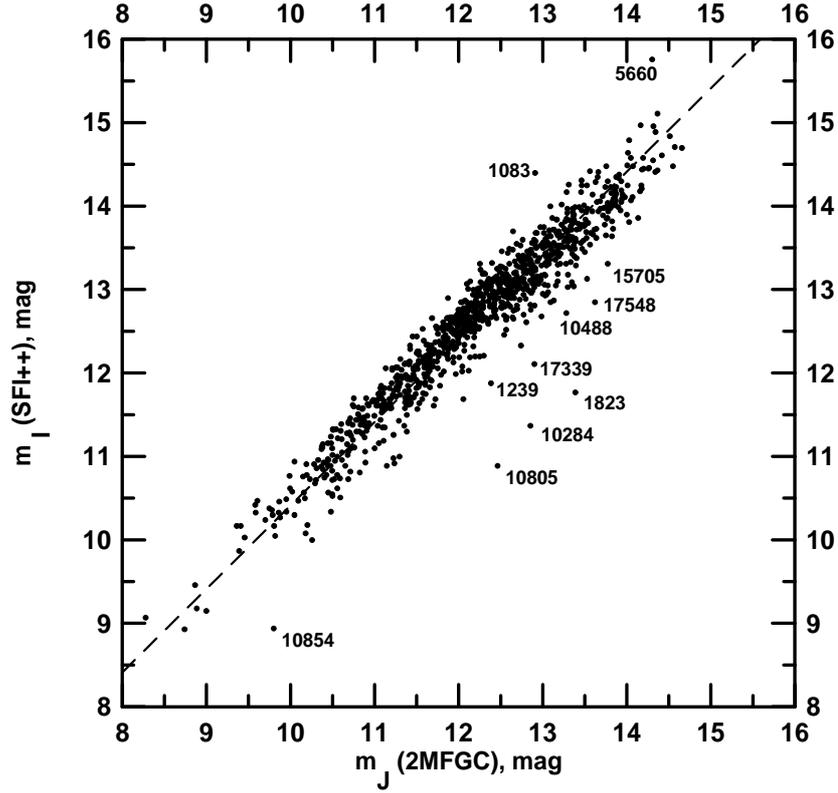

Figure 4: Comparison of corrected I-band and J-band magnitudes for 983 common galaxies in the considered samples. Deviating galaxies are denoted by their 2MFGC numbers.

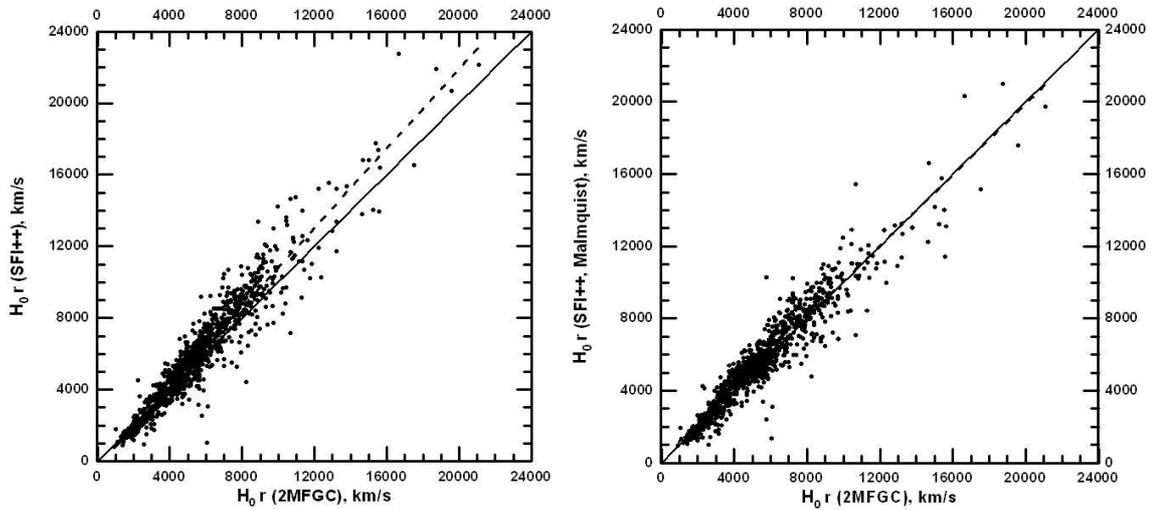

Figure 5: Comparison of distances to 983 SFI++ galaxies calculated without (left panel) and with (right panel) allowance made for the Malmquist bias with distances to the same galaxies calculated for the 2MFGC sample.

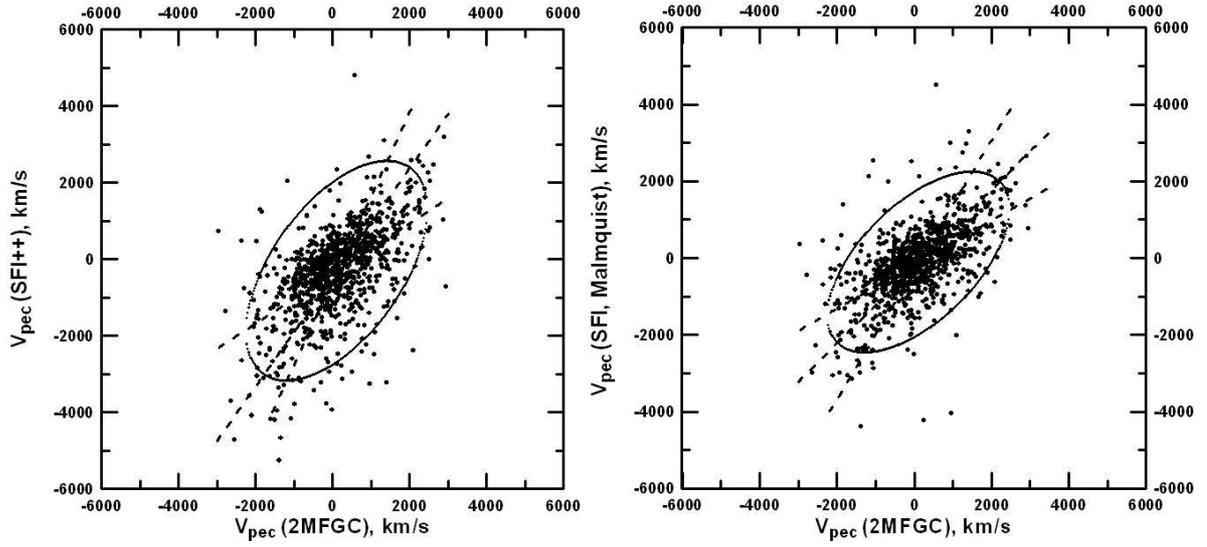

Figure 6: Peculiar velocities obtained from the SFI++ sample without (left panel) and with (right panel) allowance made for the Malmquist bias versus peculiar velocities of the same 983 galaxies derived from the 2MFGC catalog.